\documentclass[fleqn,usenatbib]{mnras}
\usepackage{graphicx}
\usepackage{natbib}
\usepackage[utf8]{inputenc}
\usepackage{multirow}
\usepackage{amsmath}
\usepackage{amsfonts}
\usepackage{amssymb}
\usepackage{ulem}
\newcommand {\uu}[1] {_{_{\rm {#1}}}}

\title{Relativistic Spherical Shocks in Expanding Media}
\author[Govreen-Segal et al.]{
    Taya Govreen-Segal\thanks{taya@govreensegal.com}, Noam Youngerman, Ishika Palit, Ehud Nakar, Amir Levinson, Omer Bromberg
	\\
	{School of Physics and Astronomy, Tel Aviv University, Tel Aviv 6997801, Israel}
}
\begin{document}
\label{firstpage}
	\maketitle		

\begin{abstract}
We investigate the propagation of spherically symmetric shocks in relativistic homologously expanding media with density distributions following a power-law profile in their Lorentz factor.  That is, $\rho_{ej} \propto t^{-3}\gamma\uu{e}(R,t)^{-\alpha}$, where $\rho_{ej}$ is the medium proper density, $\gamma\uu{e}$ is its Lorentz factor, $\alpha>0$ is constant and $t$, $R$ are the time and radius from the center. We find that the shocks behavior can be characterized by their proper velocity, $U'=\Gamma_s'\beta_s'$, where $\Gamma_s'$ is the shock Lorentz factor as measured in the immediate upstream frame and $\beta_s'$ is the corresponding 3-velocity. While generally, we do not expect the shock evolution to be self-similar, for every $\alpha>0$ we find a critical value $U'_c$ for which a self-similar solution with constant $U'$ exists. We then use numerical simulations to investigate the behavior of general shocks. We find that shocks with $U'>U'_c$  have a monotonously growing $U'$, while those with $U'<U'_c$ have a decreasing $U'$ and will eventually die out. Finally, we present an analytic approximation, based on our numerical results, for the  evolution of general shocks in the regime where $U'$ is ultra-relativistic. 
\end{abstract}

\section{Introduction}
The strong explosion problem, consisting of a sudden release of energy that drives a blast wave into the surrounding medium has been studied extensively. Special attention has been given to shocks propagating in media with power-law density profiles, where self-similar solutions can be identified. 
While in most astrophysical scenarios blast waves propagate in stationary media, there are cases in which we expect to find blast waves in expanding ones. These types of shocks can be formed when a source of energy launches the shock into a medium that was energized by an earlier explosion, setting it in motion. For example, such a scenario may arise in a binary neutron star merger, where mass is ejected during the coalescence phase, known as the \textit{dynamical ejecta} and the compact merger-product injects an additional energy into the ejecta as it accretes the remaining material. Similarly, in various types of supernovae a central engine may form following the initial explosion
injecting its energy into the expanding envelope. Indication for the existences of such engines can be found, for example, in cases where the energy measured in supernovae of type IC exceeds the amount of energy that can be formed by Nickel decay \citep{Afsariardchi2021}.

In neutron star mergers, the dynamical ejecta is likely to include a fast precursor that reaches mildly or even ultra-relativistic velocities. Mildly relativistic components were observed in numerical simulations
\citep{Bauswein2013, Hotokezaka2018, Radice2018, Ishii2018, Hotokezaka2018} though the maximal velocities, masses, and density profiles were hard to determine due to limited resolution. 
Analytic considerations suggest the possibility of an ultra-relativistic precursor \citep{Kyutoku2014,Beloborodov2020}. For example, in a model explored by \cite{Beloborodov2020} the precursor can reach high Lorentz factors with a power-law mass distribution $\rho \propto \gamma^{-\alpha}$, where $2\lesssim\alpha\lesssim4$. Finally, a mildly relativistic precursor has been suggested as an explanation for the gamma-rays observed in GW170817. The gamma-rays in this model originate from a mildely relativistic shock, 
driven by a relativistic jet that is launched following the merger, and breaks out from the ejecta
\citep{Kasliwal2017,Gottlieb2018,Nakar2020,Beloborodov2020}.
While in the case of a merger, the density profile of the fast outflows is hard to constrain, we can consider instead the density profile created as a result of a shock passing through the edge of a star. \cite{Duran2015} found (based on shock propagation theory of \citealt{Johnson1971} and \citealt{Pan2006}) that the density profile of a product ejecta formed by the passage of a mildly relativistic shock through the stellar edge
is $\rho \propto\gamma^{-\alpha}$, with $\alpha\simeq 1.6$, where $\rho$ is the proper density, and $\gamma$ is the ejecta Lorentz factor.

Motivated by the case of neutron-stars merger, where both the jet driven shock as well as the relativistic ejecta have angular dependence profiles, we consider here the simpler problem of the propagation of a spherical shock in an expanding medium with a 
power-law density profile in the Lorentz factor.
Namely, the outflow is cold and ballistic with a proper density distribution $\rho \propto \gamma\uu{e}^{-\alpha},\alpha>0$ where $\gamma\uu{e}$ is the ejecta Lorentz factor and $\alpha>0$ is chosen so that the ejecta energy is convergent.  
Though our work focuses on spherically symmetric systems, it may be applicable also to systems where over a limited angular range the symmetry is nearly spherical, such as relativistic uncollimated jets, and quasi-spherical shocks (as in the case of a shock driven by the cocoon of a choked jet). This is because relativistic blast waves are causally connected over an angular scale of $\sim \frac{1}{\gamma}$, so flows that are approximately spherical over a larger angular scale will evolve roughly as part of a spherical blast wave. 

Most previous works on relativistic shocks focused on shocks propagating in a static medium, in which case a self-similar solution can be obtained. These include \cite{Blandford1976, Sari2006} to list a few. One exception is \cite{Lyutikov2017}, who considered a double shock system, i.e.; an initial shock wave propagating in a power-law density profile followed by an additional shock or wind, and found an approximate self-similar solution for the second shock structure. While the problem bears some resemblance, it is different in the setup of the ambient medium. \cite{Lyutikov2017} considered the second shock to be very close to the first one, thus it propagates in a downstream density profile that is described by the \cite{Blandford1976} solution. We, on the other hand, consider a shock propagating in a cold, homologously expanding medium with a power-law density profile.

\cite{Govreen-Segal2021} studied the Newtonian analogous to the setup considered here. In their setup the shock was propagating in a homologously expanding medium with a power-law density profile: $\rho\propto v^{-\alpha}$. They found that for profiles with $\alpha\lesssim8$, all shocks decay, i.e. the ratio between the shock velocity and the immediate upstream velocity decreases with time. In steeper density profiles, for every density profile, they found a critical ratio  between the shock velocity and immediate upstream velocity, such that in shocks with a ratio larger than the critical value, the ratio grows monotonically with time, while shocks with a smaller ratio, monotonically decay. Separating these two regimes is a self-similar solution, describing a shock with a constant ratio between the shock and the immediate upstream velocities, equal to the critical value.

Similar to the Newtonian case \citep{Govreen-Segal2021}, since there are two velocity scales, we generally do not expect to find a self-similar solution. However, we find that for every density profile with $\alpha>0$, there exists a critical ratio between the shock Lorentz factor and the immediate upstream Lorentz factor for which a self-similar solution exists. We then use numerical simulations to study the evolution of general shocks, which are not self-similar. 

We proceed as follows: in \S\ref{sec: expanding} we define the setup we're considering and give an overview of the solutions and go on to derive the self-similar solutions in \S\ref{sec: self-similar}. We then use numerical  simulations to study general solutions in \S \ref{sec: Numerical}. Finally, in \S\ref{sec: Summary} we conclude.

\section{Shock propagation in a homologously expanding medium}
\label{sec: expanding}
Consider a spherically symmetric, relativistic shock wave that propagates through a cold, expanding medium. The expansion of the medium is assumed to be homologous, with a velocity $v\uu{ej}$ related to the radius $r$ and time $t$ by the equation $v\uu{ej}=r/t$ for $r<ct$, where $c$ is the speed of light. We model the density profile of the medium as a power law 

\begin{equation}
\rho\uu{ej} =a\uu{ej} t^{-3}\gamma\uu{ej}^{-\alpha},
\label{density}
\end{equation}
where $\rho\uu{ej}$ is the proper density, $a\uu{ej}$ is a constant, and $\gamma\uu{ej}=\gamma\uu{ej}(r,t)$ is the Lorentz factor of the ejecta, which we assume to be ultra-relativistic. We focus on cases with $\alpha>0$, which correspond to a scenario where most of the ejecta energy is stored in slow material. We assume 
a relativistic ideal equation-of-state with an adiabatic index $\hat{\gamma} = 4/3$.
Our goal is to characterize the evolution of a spherical blast wave in such a medium. Specifically, we aim to find under which conditions such a shock decays and ultimately dies out and under which conditions it grows and crosses an infinite number of mass shells. 

We denote by capital letters shock-related properties and by sub-index $e$ the medium properties at the location of the shock front, i.e. at the immediate upstream, measured in the lab frame. Namely, $R, V, U, \Gamma$ represent
the shock radius, absolute values of the 3-velocity and 4-velocity and the Lorentz factor respectively, while $v\uu{e}\equiv\frac{R}{t}$ and $\gamma\uu{e}$ represent the medium 3-velocity and Lorentz factor at the immediate upstream respectively. Untagged quantities are measured in the lab, or in the proper frame, according to the regular convention. In addition we denote by capital-tagged letters shock properties measured in the frame of the \textit{immediate upstream}\footnote{Since the upstream is moving, its proper frame is different than the lab frame.}.

Let us examine which parameters may affect the shock evolution. Once the initial conditions are forgotten, in addition to the adiabatic index $\hat{\gamma}$ and the density power-law index $\alpha$, the shock evolution can only depend on the shock radius $R$, the shock velocity $V$, the medium density and velocity at the immediate upstream $\rho\uu{e}, v\uu{e}$ and on time. 
 Since all parameters evolve as powerlaws, $a\uu{e}$ cannot be a relevant parameter. In addition, since $R$ and $t$ are connected via the ratio $\frac{R}{t}$, which by definition is $v\uu{e}$, we are left with two parameters $\{v\uu{e},V\}$. Finally, since all velocities are relativistic we replace the 3-velocities with the corresponding Lorentz factors and obtain 
 that the shock evolution depends only on $\hat{\gamma},\alpha$ and on the ratio $\frac{\Gamma}{\gamma\uu{e}}$, or equivalently on $U'\simeq\frac{(\Gamma/\gamma\uu{e})^2-1}{2\Gamma/\gamma\uu{e}}$, the shock 4-velocity in the immediate upstream frame, appropriate for a case where $\Gamma,\gamma\uu{e}\gg1$.
 Note that while we assume that the shock and the immediate upstream are ultra-relativistic in the lab frame, the shock may be mildly relativistic or even Newtonian in the upstream frame.

 As stated above, the fact that the flow has two velocity scales implies that we generally do not expect the solution to be self-similar. An exception is if $U'$ is constant throughout the shock evolution. In such a case, there may be a self-similar solution, i.e.; shocks with a critical value of
 $U_c'(\alpha)=Const$. For density profiles in which such a solution exists, $U'$ will either monotonically increase or monotonically decrease, depending on whether the value of $U'$ is larger or smaller than the self-similar value, thus defining two qualitatively different types of regimes. The first, termed \textit{growing shocks}, for which $U'$ increases monotonically with time  asymptotically approaching infinity, 
 and a second type of \textit{decaying shocks}, consisting of shocks with $U'$ that decreases with time. In all decaying shock the shock Lorentz factor ultimately approaches the local Lorentz factor of the moving medium, viz., $\Gamma\to\gamma\uu{e}, U'\to 0$, where the shock dies out. This condition differs from the static case, in which a decaying shock decelerates, and its velocity approaches zero in the lab frame. It is worth noting that as viewed in the lab frame, both decaying and growing shocks accelerate. 

The shock behaviour in the two regimes, i.e.
shocks with $U'>U'_c$ grow while shocks with $U'<U'_c$ decay, renders the self similar solution a repelling one (a bifurcation point of shock solutions). In addition, as the energy in the ejecta increases for smaller $\alpha$, becoming infinite at $\alpha<0$, we expect a minimal value of $\alpha$ below which no self-similar solution exists. i.e. $U_c'\underset{\alpha\to \alpha_{min}}{\to} \infty$. 
Below, we derive a self-similar solution with a constant $U'$, and find that such a solution exists for every $\alpha>0$.

\subsection{Self-similar solutions for $U'=U_c'$}
\label{sec: self-similar}
We seek a self-similar solution where for a given value of $\alpha$, $U'=U_c'$, or equivalently $\Gamma/\gamma\uu{e}=Const$. 
The requirement that $\Gamma/\gamma\uu{e}=Const$ obeys the scaling  $\Gamma^2(t) = At^m$, where $m$ is a free parameter and $A$ is a scaling constant. This can be shown through the following argument.
Denoting the shock velocity as $V$, the shock trajectory is given to an order O($\Gamma^{-2}$) by 
\begin{align}\label{eq: Rs}
\begin{split}
R(t) = \int_0^t V(t')dt' &= \int_0^t{\left(1-\frac{1}{2\Gamma^2}\right){\rm d}t^\prime}\\&=t-\frac{t}{2(1- m)\Gamma^2}.
\end{split}
\end{align}
Using Eq. \eqref{eq: Rs} we can obtain the medium velocity at the shock immediate upstream, 
\begin{equation}
    v\uu{e}=R/t=1-1/[2(1-m)\Gamma^2],
\end{equation}
and its Lorentz factor (to an order O($\Gamma^{-2}$)):
\begin{equation}
\gamma\uu{e}^2=(1-m)\Gamma^2,
\label{g-ejecta}
\end{equation} 
implying that $\gamma\uu{e}/\Gamma=\sqrt{1-m}$ and 
$U_c'=\frac{m}{2 \sqrt{1-m}}$. 
The density at the shock location is given by
\begin{equation}
\rho\uu{e}=\frac{a\uu{e}}{t^3\gamma\uu{e}^\alpha}=\frac{a\uu{e} A^{3/m}}{(1-m)^{\alpha/2}}\Gamma^{-(\alpha+6/m)}.
\end{equation}
From (\ref{g-ejecta}) it is seen that $\gamma\uu{e} < \Gamma$ implies  $0 < m <1$. 
A convenient choice of a similarity parameter is
\begin{equation}
\sigma =[1+2(1-m)\Gamma^2](1-r/t),\label{chi}
\end{equation}
such that the shock is located at
$\sigma(R)=1+\frac{1}{2\gamma\uu{e}^2}\rightarrow1$
to the order we are working with here.
Using the similarity parameter, we may re-write the shock downstream parameters $\gamma,p,\rho$ in terms of the self-similar parameters and reduce the equations of relativistic fluid dynamics to ordinary differential equations, which can be easily solved. 
The solution requires boundary conditions, attained by the shock jump conditions. 

We denote the variables of the shocked medium with subscript $2$ ($\gamma_2, \rho_2$, etc). The downstream enthalpy is given by $w_2=\rho_2h_2\gamma_2^2$, where $h_2 = 1+\hat{\gamma} p_2/[\rho_2(\hat{\gamma}-1)]$ is the enthalpy per baryon. 
The jump conditions at the shock ($\sigma=1$) read:
\begin{align}
\rho\uu{e}\gamma\uu{e}(v\uu{e}-V)&=\rho_2\gamma_2(v_2-V),\label{jump-rho}\\
\rho\uu{e}\gamma^2_e(v\uu{e}-V)&=w_2(v_2-V)+p_2V,\label{jump-mom}\\
\rho\uu{e}\gamma^2_e v\uu{e}(v\uu{e}-V)&=w_2v_2(v_2-V)+p_2.\label{jump-energ}
\end{align}
Equations (\ref{jump-rho})-(\ref{jump-energ}) can be solved by employing Eq. (\ref{g-ejecta}):
\begin{align}
\gamma^2_{2}(1)&=q\Gamma^2, \\
\rho_{2}(1)\gamma_2(1)&=\frac{mq \rho\uu{e}\gamma\uu{e}}{(1 - m)(1 - q)},\\
p_{2}(1)&=\frac{m \rho\uu{e}}{2- \frac{\hat{\gamma}}{\hat{\gamma}-1}(1- q) }\left(\sqrt{\frac{q}{(1- m)}} -1 \right),
\end{align}
where $\sqrt{q}$ is the only positive solution of the equation
\begin{equation}
\hat{\gamma}x^3+(2-\hat{\gamma})\sqrt{1 - m} \, x^2-(2-\hat{\gamma})x-\hat{\gamma}\sqrt{1-m} =0.\label{q}
\end{equation}
Note that $q=1$ for $m=0$ and $q=1/2$ for $m=1$.
We can now define the self-similar variables to be:
\begin{align}
    G(\sigma) &= \gamma^2_2/\gamma^2_2(1)\\
    F(\sigma) &= p_2/p_2(1)\\
H(\sigma) &=\rho_2\gamma_2/(\rho_2(1)\gamma_2(1)).
\end{align}
With these definitions, the fluid equations in the shocked region reduce to:
\begin{eqnarray}
2(1+qG\sigma)\frac{d\ln F}{d\sigma}-(1-qG\sigma)\kappa\frac{d\ln G}{d\sigma}=\nonumber\\
\frac{(\kappa \, m - mn -6)}{(1- m)}qG,\label{in1}\\
2(1-qG\sigma)\frac{d\ln F}{d\sigma}-\hat{\gamma}(1+qG\sigma)\frac{d\ln G}{d\sigma}=\nonumber\\
\frac{[ \hat{\gamma}(2 - m) + mn-6(\hat{\gamma}-1)]}{(1 - m)}qG,\label{in2}\\
2(1-qG\sigma)\frac{d\ln H}{d\sigma}-2\frac{d\ln G}{d\sigma}=\nonumber\\-\frac{(m - m n -2)}{(1 - m)}qG,\label{in3}
\end{eqnarray}
where $\kappa(\sigma)=w_2/p_2\gamma_2^2$.  The boundary conditions are $G(1)=F(1)=H(1) =1$.
The first two equations can be expressed 
as 
\begin{eqnarray}
\frac{d\ln F}{d\sigma} = \frac{\Delta_1}{\Delta}, \qquad 
\frac{d\ln G}{d\sigma} = \frac{\Delta_2}{\Delta},
\end{eqnarray}
where 
\begin{align}
\Delta &= c_s^2-\frac{(G q \sigma -1)^2}{(G q \sigma +1)^2} \label{D}, \\ 
\Delta_1 &= \frac{Gq(\kappa\left(\alpha m+6\right)\left(Gq\sigma-1\right)}{2\kappa\left(1-m\right)\left(Gq\sigma+1\right)^{2}} - \label{D1}  \\ &\hat{\gamma}\left(m (\alpha +\alpha  G q \sigma -2 \kappa )+4 \kappa  (G q \sigma -1)+6 G q \sigma +6\right)),\nonumber
\\
\Delta_2 &= \frac{Gq}{\kappa\left(1-m\right)\left(Gq\sigma+1\right){}^{2}} \cdot \label{D2} \\ & \left(m\left(Gq\sigma\left(\hat{\gamma}-\kappa\right)+\kappa-2\alpha+\hat{\gamma}\right)+4\left(\hat{\gamma}+\hat{\gamma}Gq\sigma-3\right)\right)\nonumber .
\end{align}
Note that the dimensionless sound speed is related to $\kappa$ through
\begin{equation}
c^2_s = \frac{\hat{\gamma} p_2}{h_2\rho_2} = \frac{\hat{\gamma}}{\kappa},
\end{equation}
and the sonic point occurs at $\Delta =0$.

\subsubsection{Characteristics}
The  velocities of the $C_\pm$ characteristics in a relativistic flow are given by:
\begin{equation}
\frac{d r_\pm}{dt} = \frac{v\pm c_s}{1\pm v c_s}.
\label{Char+}
\end{equation}
Now, define $\sigma_+(t)$ to be the value of the self-similar coordinate of the  $C_+$ characteristic.  We have  
\begin{equation}
\frac{d\sigma_+}{dt} = 2(1-m)t^{-1}\Gamma^2[m(1-r_+/t) + r_+/t -dr_+/dt].
\end{equation}
Substituting Eq. (\ref{Char+}) we obtain

\begin{align}
    \frac{d\sigma_+}{dt} &= 2(1-m)t^{-1}\Gamma^2 \left[\frac{(1-v)(1-c_s)}{1+vc_s} - \frac{\sigma}{2\Gamma^2}  \right] \nonumber\\
&= \frac{1-m}{t qG}\left[\frac{1-c_s}{1+c_s} - qG\sigma\right].
\end{align}
It is seen that for the particular characteristic for which  $\Delta =0$,  the term in the parentheses vanishes and $d\sigma_+ /dt =0$.
That is, the sonic point is located at a fixed self-similar coordinate, as in the Newtonian case. 

\subsubsection{The Sonic Point}
Since at the sonic point $\Delta=0$, a smooth crossing of this point requires that $\Delta_1=\Delta_2=0$ at that point as well. This, in turn, fixes $ m $ for every value of $ \alpha $.
In order to determine the eigenvalue $m$ for a given setup with known values of $\alpha$ and $\hat{\gamma}$, we 
integrate Eqs. \eqref{in1}-\eqref{in3} numerically together with the boundary conditions on the shock front,
and search for a value of "m" that allows for a smooth transition through the sonic point. 

\begin{figure}
	\center
	\includegraphics[width=\columnwidth]{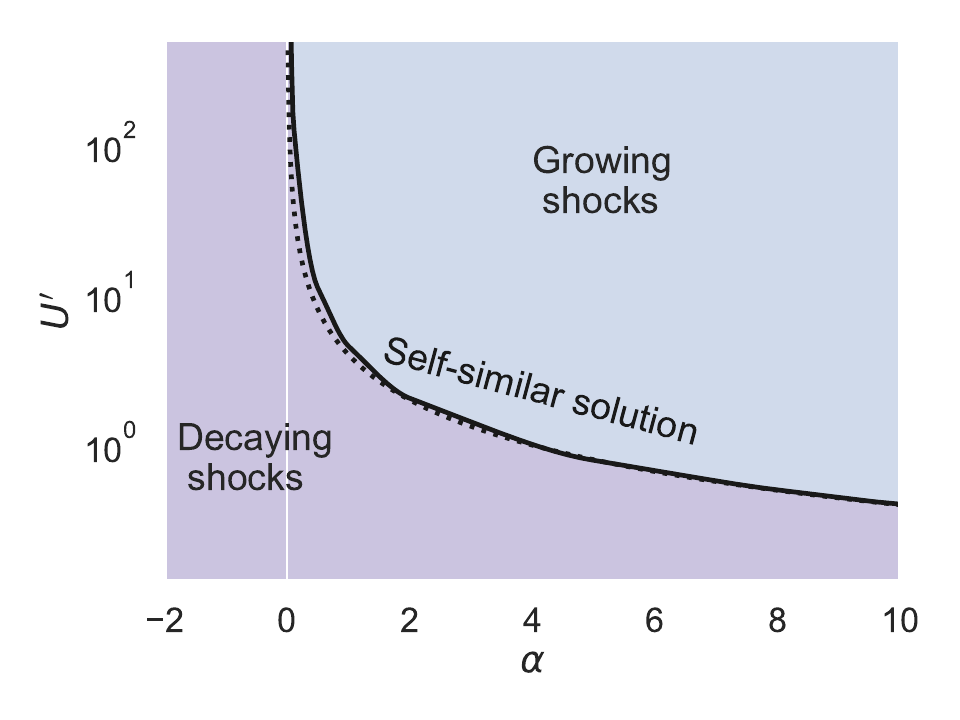}
	\caption{The phase space of the analytic solution. The black line shows the dependence of the self-similar value of $U'$ ($U'_c=\dfrac{m}{2\sqrt{1-m}}$) on $\alpha$, found by requiring a smooth transition through the sonic point. The white line at $\alpha=0$ marks the asymptote at which the self-similar solution diverges and terminates. The black dotted line depicts the relation $U'=4/\alpha$, which is in good agreement with the semi-analytic solution for $U'_c$. The self-similar solution divides the parameter space into two distinct regions; growing shocks, and decaying shocks.}
	\label{fig:U_tag(alpha)}
\end{figure}

Figure \ref{fig:U_tag(alpha)} displays the relationship between $U_c'$ and $\alpha$ for the family of self-similar solutions. We find that a solution exists for cases with $\alpha>0$. As expected, $U_c'$ approaches infinity ($m\to 1$) as $\alpha\to 0$, and converges to zero as $\alpha$ becomes large. We find that to a very good approximation, $U'_c=\frac{4}{\alpha}$ in the range tested above.
The drop in $U_c'$ is notably abrupt, with a value of $\frac{\Gamma}{\gamma\uu{e}} \approx 1.5$ at $\alpha=5$.  

\section{Numerical Results}
\label{sec: Numerical}
To study the shock evolution in the general, non-self-similar cases, we use numerical simulations.

\subsection{Simulation Setup}
\label{sec: Simulation Setup}
We use the publicly available code GAMMA\footnote{https://github.com/eliotayache/GAMMA} \citep{Ayache2022} 
to carry out 1D relativistic hydrodynamic (RHD), spherically symmetric simulations. We set up an initial blastwave that propagates in an expanding medium and follow its evolution to times when it is no longer affected by the initial conditions.  Using GAMMA, we are able to properly resolve shocks up to a Lorentz factor of 400. 

The initial conditions are set such that the upstream has a density profile of $\rho\uu{e}\propto(\gamma\uu{e} v\uu{e})^{-\alpha}$ and a velocity profile $v\propto r$. 
As the simulation starts, the shock, which was present in the computational domain begins to propagate in the medium. Once the initial conditions are forgotten, 
the shock location, time, immediate downstream Lorentz factor, and immediate upstream Lorentz factor are collected from each simulation. The simulations are then grouped according to the value of $\alpha$. Note that for every value of $\alpha$, there are several simulations ranging in different values of $\Gamma/\gamma\uu{e}(R)$. The fact that in Fig \ref{fig:phase_space} (which will be discussed in the following subsection) the different simulations form a continuous curve in the $\frac{\Gamma}{\gamma\uu{e}(R)}-\frac{\Gamma/\gamma\uu{e}(R)}{d\log\gamma\uu{e}(R)}$ space, shows that the initial conditions are indeed forgotten and that the simulation is at a high enough resolution to accurately resolve the shock Lorentz factor. 
The simulation details and initial setup are discussed in more detail in appendix \ref{appendix: numerical}.

\subsection{Simulation Results}
\label{sec: results}
\begin{figure}
    \centering
    \includegraphics[width=\columnwidth]{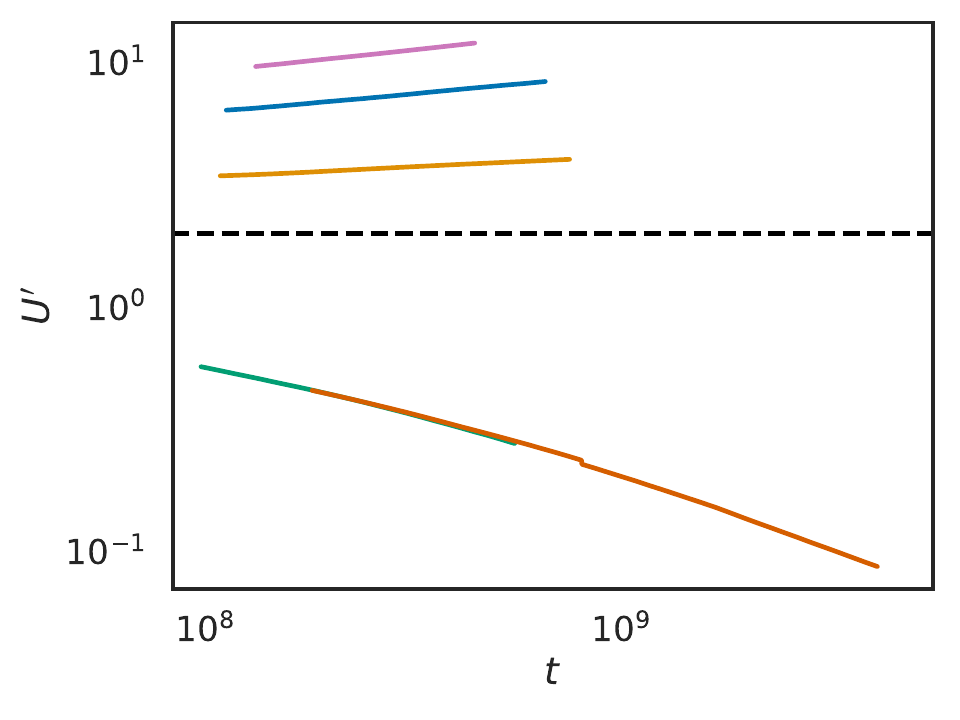}
    \caption{The evolution of the shock 4-velocity measured in the immediate upstream frame, $U'(t)$, in several simulations with a medium having a powerlaw density profile with $\alpha=2$. The date from each simulation is marked with a different color. The black dashed line marks self-similar value $U'_c$. Time is measured in arbitrary units.}
    \label{fig:alpha2_sims}
\end{figure}

Fig. \ref{fig:alpha2_sims} shows $U'(t)$ for several simulations in a density profile with $\alpha=2$. 
The simulations are plotted from the time where the shock reaches a self-consistent structure, independent of the initial conditions. Each simulation is marked with a different color, and the dashed line mark the value of $U'_c$. As expected, shocks with an initial $U'<U'_c$ ($U'>U'_c$) have a monotonically increasing (decreasing) $U'(t)$ throughout the entire simulation.   

\begin{figure}
\centering
\includegraphics[width=\columnwidth]{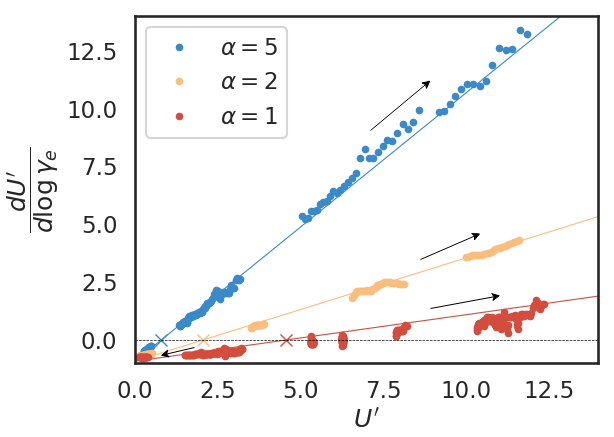}
\caption{The evolution of the shock strength defined by the parameter $\frac{dU'}{d\log\gamma_e}$, which measures the change in $U'$ as it crosses the upstream 
as a function of $U'$, is plotted for different values of the density profile. The figure shows several simulations for every value of $\alpha$. The values from the numerical simulations are marked by dots, and the self-similar solutions all lie on the dotted black line at $(U'_c\left(\alpha\right),0)$ and are marked with 'x'. Regions above the dotted line correspond to growing shocks, while the ones below it are decaying shocks. The solid lines denote the analytical approximation (see \S\ref{sec: approx}).}
	\label{fig:phase_space}
\end{figure}

Fig. \ref{fig:phase_space} shows the different regimes of shock evolution for different values of $\alpha$. Each point represent a snapshot from a simulation, where all simulations with the same $\alpha$ are given the same color, according to the color legend. The self-similar solution is marked with an $x$ colored according to the corresponding value of $\alpha$. It sits on the $\frac{dU'}{d\log\gamma_e}=0$ line 
(shown with a black dotted line), which divides the parameter plane into two regimes:  The domain below the line corresponds to shocks that decay with time. In this case $\frac{dU'}{d\log\gamma_e}<0$ and the trajectory evolution of the shock is downwards and to the left (shown with a black arrow). 
In the region above the self-similar line, $\frac{dU'}{d\log\gamma_e}>0$ and shocks move upwards and to the right with time as they grow. 
For $\alpha=1$ (red dots), we find that the simulations are in slight disagreement with the self-similar solution, and do not pass through it exactly, but rather are slightly below it. The reason is likely numerical. For 
smaller values of $\alpha$, it takes the simulation longer to forget the initial conditions, and within the limited dynamical range allowed by our computational resources it is likely that the shock is still affected by the initial conditions.

\subsection{Analytic Approximation} \label{sec: approx}
Examining Fig. \ref{fig:U_tag(alpha)}, we notice that at least for $U'\gg1$, $\frac{dU'}{d\log\gamma\uu{e}}$ appears to be linear with respect to $U'$. As this approximation must pass through $(U'_c,0)$, our ansatz takes the form:
\begin{equation}
\frac{dU'}{d\log \gamma\uu{e}}= a\left(\frac{U'}{U'_c}-1\right),
\end{equation}
where $-a$ is the intercept. Measuring $a$ from the numerical simulations in this manner is not robust, as the numerical differentiation introduces a lot of noise. We therefore first integrate the relation to find:
\begin{equation}\label{eq:U'}
    U'_c\log\left|U'-U'_c\right|=a\log\gamma\uu{e}+c
\end{equation}
where $c$ is an arbitrary constant that depends on the initial conditions. We can use this expression to fit a linear relation and find $a$. In our simulations, we find that $a\simeq 0.75-0.8$ for $U'\simeq 0-2$, and $a\simeq 0.9-0.95$ for $U'\gg1$. In Fig. \ref{fig:U_tag(alpha)}, we adopt $a=0.92$. The analytic approximation is shown in Fig. 
\ref{fig:phase_space} with thin solid lines.

For more general use, it may be useful to re-write \eqref{eq:U'} as:
\begin{equation}
    U'= (U_0' - U_c')\left(\frac{\gamma\uu{e}}{\gamma\uu{e,0}}\right)^{a/U'_c}+U'_c
\end{equation}
where $U'_0$ and $\gamma\uu{e,0}$ are the initial values taken at a time when the simulation has evolved to a point where the shock evolution becomes independent of the initial conditions. Note that for $U_0'> U'_c$ \, ($U_0' < U'_c$) the shock is indeed growing (decaying).

\section{Summary}
\label{sec: Summary}
In this paper, we study the propagation of a spherically symmetric shock in a relativistic homologously expanding medium with a power-law density gradient ($\rho_{ej} \propto t^{-3} \gamma\uu{e}\left(R,t\right)^{-\alpha}$, $\alpha>0$). The medium is assumed to consist of an ideal cold gas where the adiabatic index of the shocked gas is $\hat{\gamma}=4/3$. This index is applicalbe for all relativistic shocks and for Newtonian shocks where the downstream internal energy is dominated by radiation. Note that while the setup is spherical, our solutions are applicable to jets and quasi-spherical shocks if they are relativistic enough so the jet core is not causally connected with the edges. 
We find that while the shock always accelerates in the lab frame, the shock behavior can be characterized according to the shock four-velocity as measured in the immediate upstream frame - $U'$. Once the initial conditions are forgotten, $U'$ may either increase monotonously, corresponding to a growing shock, or decrease, meaning the shock is decaying throughout the evolution. Separating these two regimes there exists an unstable self-similar solution, for which $U'$ is constant. That is, for every $\alpha>0$ there is a critical value $U'=U_c'$ above which the shock grows and below which it decays.

The self-similar value diverges as $\alpha\to 0$, and decreases with $\alpha$, so that for $\alpha\gtrsim 5$, in the self-similar case, $U'$ is mildly-relativistic or even Newtonian. 
We present an analytical approximation that can be used to describe general shocks, and seems to be robust as long as $U'$ is ultra-relativistic.

\section*{Acknowledgements}
This research was partially supported by a consolidator ERC grant 818899 (JetNS) and by an ISF grant (1995/21). 
AL acknowledges support by a grant from the Simons Foundation (MP-SCMPS-00001470). TGS thanks the Buchman Foundation for their support. OB and IP acknowledge the support of an ISF grant 1657/18, a BSF grant 2018312 and an NSF-BSF grant 2020747.

\section*{Data Availability}
The data underlying this article will be shared on reasonable request to the corresponding author.

\bibliographystyle{mnras}
\bibliography{spherical_shocks}%

\appendix
\section{More details on the Numerical Simulations}\label{appendix: numerical}
\subsection{Simulation Setup}
In all our simulations with GAMMA we use piece-wise linear spacial reconstruction, hllc solver and third order Runge-Kutta time stepping and a CFL of 0.4. 

The grid initially spans from $R_{0}\cdot(1-10/\Gamma^2)$, till where the Lorentz factor of the upstream reaches 50,  where $R_0$ is the initial shock radius, and $\Gamma$ is the initial shock Lorentz factor in the lab frame. The inner boundary is reflective, and is stationary throughout the simulation, while the outer boundary is set to outflow and moves at $1.05c$ ($c$ is the speed of light), forming a region ahead of the initially set density profile with uniform density pressure and Lorentz factor. The simulation is stopped before the shock enters this region. 

While the grid resolution is initially uniform, with 5000 cells, within a few time steps the AMR re-sets the resolution. The re-gridding scheme is set to run away in order to fully resolve the shock. We set the maximum number of cells to 20,000, which we find is high enough to never be necessary.  We set the re-gridding score as $S_{regrid}=\frac{\Delta r}{r_{max} \Delta \theta}\gamma^{3/2}$, where $\Delta \theta = \frac{\pi}{6000}$, and allow $S_{regrid}$ to vary in $[0.1,3]$, the resolution in the 10 cells ahead of the shock is increased by a factor of 10. 
\subsection{Convergence}
One way to verify the convergence of the simulation, and make sure that the simulation correctly captures the shock is to derive the shock Lorentz factor in two ways and compare them. The first, is by following the shock location $R(t)$, and calculating the Lorentz factor that corresponds to the velocity $V=\frac{dR}{dt}$.
The second way is to measure the Lorentz factor of the material at the immediate downstream an upstream of the shock and use the shock jump conditions to obtain the shock Lorentz factor. 
Comparing the two, we find a perfect agreement. In fact, plotting \ref{fig:U_tag(alpha)} in both methods results in identical plots. 
\end{document}